**Astro2020 Science White Paper**

# Cosmic Dawn and Reionization: Astrophysics in the Final Frontier

**Thematic Areas:** ☐ Planetary Systems ☐ Star and Planet Formation
☐ Formation and Evolution of Compact Objects ☐ Cosmology and Fundamental Physics
☐ Stars and Stellar Evolution ☐ Resolved Stellar Populations and their Environments
☒ Galaxy Evolution ☐ Multi-Messenger Astronomy and Astrophysics

**Principal Author:**
Name: Asantha Cooray
Institution: University of California, Irvine
Email: acooray@uci.edu
Phone: 949 824 6832

**Co-authors:** (names and institutions)
James Aguirre (Penn), Yacine Ali-Haimoud (NYU), Marcelo Alvarez (Berkeley), Phil Appleton (Caltech), Lee Armus (Caltech), George Becker (Riverside), Jamie Bock (Caltech/JPL), Rebecca Bowler (Oxford), Judd Bowman (ASU), Matt Bradford (JPL), Patrick Breysse (CITA), Volker Bromm (Austin), Jack Burns (Colorado), Karina Caputi (Kapteyn), Marco Castellano (INAF), Tzu-Ching Chang (JPL), Ranga Chary (Caltech), Hsin Chiang (McGill), Joanne Cohn (Berkeley), Chris Conselice (Nottingham), Jean-Gabriel Cuby (LAM), Frederick Davies (UCSB), Pratika Dayal (Kapteyn), Olivier Dore (JPL), Duncan Farrah (Hawaii), Andrea Ferrara (Pisa), Steven Finkelstein (UT Austin), Steven Furlanetto (UCLA), Bryna Hazelton (UW), Caroline Heneka (SNS), Anne Hutter (Kapteyn), Daniel Jacobs (ASU), Leon Koopmans (Kapteyn), Ely Kovetz (Ben-Gurion), Paul La Piante (Penn), Olivier Le Fevre (LAM), Adrian Liu (McGill), Jingzhe Ma (UCI), Yin-Zhe Ma (KawZulu-Natal), Sangeeta Malhotra (Goddard), Yi Mao (Tsinghua), Dan Marrone (Arizona), Kiyoshi Masui (MIT), Matthew McQuinn (Washington), Jordan Mirocha (McGill), Daniel Mortlock (Imperial), Eric Murphy (NRAO), Hooshang Nayyeri (UCI), Priya Natarajan (Yale), Thyagarajan Nithyanandan (NRAO), Aaron Parsons (Berkeley), Roser Pello (OMP), Alexandra Pope (UMass), James Rhoads (Goddard), Jason Rhodes (JPL), Dominik Riechers (Cornell), Brant Robertson (UCSC, IAS), Claudia Scarlata (Minnesota), Stephen Serjeant (Open U), Benjamin Saliwanchik (Yale), Ruben Salvaterra (INAF), Rafffaella Schneider (Rome-La Sapienza), Marta Silva (Oslo), Martin Sahlén (Uppsala), Mario G. Santos (Western Cape), Eric Switzer (Goddard), Pasquale Temi (Ames), Hy Trac (CMU), Aparna Venkatesan (San Fransisco), Eli Visbal (Flatiron), Matias Zaldarriaga (IAS), Michael Zemcov (RIT), Zheng Zheng (Utah)

**Abstract** (optional):
   The cosmic dawn and epoch of reionization mark the time period in the universe when stars, galaxies, and blackhole seeds first formed and the intergalactic medium changed from neutral to an ionized one. Despite substantial progress with multi-wavelength observations, astrophysical process during this time period remain some of the least understood with large uncertainties on our existing models of galaxy, blackhole, and structure formation. This white paper outlines the current state of knowledge and anticipated scientific outcomes with ground and space-based astronomical facilities in the 2020s. We then propose a number of scientific goals and objectives for new facilities in late 2020s to mid 2030s that will lead to definitive measurements of key astrophysical processes in the epoch of reionization and cosmic dawn.

**I. Cosmic Dawn and Reionization are ripe for discoveries.** Cosmic Dawn – when the first stars lit up the Universe, eventually coalescing into the first generations of galaxies – and the epoch of reionization (EoR) that follows are among the most exciting frontiers in cosmology and astrophysics (Barkana & Loeb 2001; Dayal & Ferrara 2008). While precise details are still missing, it is now believed that sometime between 200 and 600 Myr after the Big Bang ($z \sim 10$–$20$), the first collapsed objects formed in dark matter halos and combined with subsequent generations of galaxies, eventually produced enough UV photons to reionize all of the surrounding hydrogen gas in the intergalactic medium (IGM). The same photons, or another process, also heated the intergalactic medium (IGM) to temperatures of order $2 \times 10^4$ K from 10-1000 K.

The era of cosmic dawn and reionization (Figure 1) is a crucial chapter in the history of galaxies, active galactic nuclei (AGN), IGM, and heavy elements in the universe. Large-angle CMB polarization suggests an optical depth to reionization of $0.054 \pm 0.007$ (Planck Collaboration 2018), of which ~0.04 arises from $z < 6$. This corresponds to a reionization redshift of $7.7 \pm 0.8$, if reionization is modeled as instantaneous. It is likely that the reionization process is extended in time and spatially inhomogeneous.

*How did the universe reionize?* Existing theoretical studies suggests that a combination of the first metal-free Population III stars (Bromm & Larson, 2004), the subsequent generations of Population II stars, and accretion onto remnant black holes such as mini-quasars and quasars (Venkatesan et al., 2001) are responsible for the UV photons that reionized the universe. Other sources of ionizing photons are decays and annihilation of light dark matter particles'sterile neutrinos (Mapelli, Ferrara & Pierpaoli 2006), primordial metal-poor globular clusters (Ricotti & Shull 2000), cosmic rays (Tueros et al. 2014), and supernovae shocks (Johnson & Khochfar 2011).

The preferred source of ionizing photons is galaxies. As the UV luminosity functions (LFs) at $z > 6$ have faint-end slopes that are steep (Bouwens et al., 2015; Finkelstein et al., 2015, Yue et al. 2018), UV photon density is inferred to be dominated by low mass, low luminosity systems (Salvaterra et al. 2011). While galaxy number counts may be adequate, one major uncertainly is the *escape fraction*, the fraction of ionizing photons that are able to escape the galaxies to ionize the surrounding neutral intergalactic medium. Existing reionization models require escape fractions at the level of 20% (Robertson et al., 2015). Such a high escape fraction is inconsistent with direct measurements, which are at the level of a few % for bright luminous galaxies (Hayes et al., 2011; Vanzella et al., 2012). This difference could be partly alleviated if the UV luminosity density inferred from $z > 6$ GRBs is used to model reionization (Chary et al. 2016). Numerical simulations suggest that fainter and smaller galaxies in low mass dark matter halos may have higher escape fractions (Xu et al. 2016), leading to a mass- or luminosity-dependent escape fraction for ionizing photons during reionization. The current uncertainties on the faint-end slope of the UV LFs at $z > 6$ and the escape fraction of galaxies are such that galaxies alone may not be responsible for reionization. Alternatively, models that do allow reionization with galaxies alone have tensions with other observables, e.g., the ionized volume filling fraction (Finkelstein et al. 2019).

Some fractional contribution of UV photons are expected from quasars and AGNs (Madau & Haardt, 2015; Willott et al. 2010). A proper understanding of the role of AGNs in reionization also require tracing back those sources to their formation pathways, allowing studies related to primordial or seed blackholes in the universe (Valiante et al. 2018). The fractional contribution from light dark matter particle decays and sterile neutrinos are more uncertain. If an astrophysical process during reionization can be accurately traced to photons from particles, then reionization observations may also become a tool for studies on fundamental physics.

*How did the universe reheat?* HI 21-cm line intensity and spatial anisotropies are currently the best identified probes of the thermal state of the IGM, though 21-cm measurements do not reveal the source of heating. Initial HI 21-cm line anisotropy power spectrum measurements, with LOFAR, MWA etc, suggest heating must have happened before $z > 9.6$ (Patil et al. 2017). The 21-cm global signature measurement with EDGES (Bowman et al. 2018) shows an absorption that could be interpreted as evidence for presence of Ly-α coupling at $z=17$ and that the IGM was heated above the CMB temperature at z of 15. It could be that the same sources responsible for ionizing photons may also be responsible for heating of the IGM. The leading candidates for heating are soft X-rays either from stellar remnants in primordial galaxies or gas accretion to primordial blackholes, or both (Oh & Zaiman 2002). Other minor possibilities involve shock heating of the IGM (Furlanetto & Loeb 2004) and resonant scattering of Ly-α photons (Madau, Meiksin & Rees 1997).

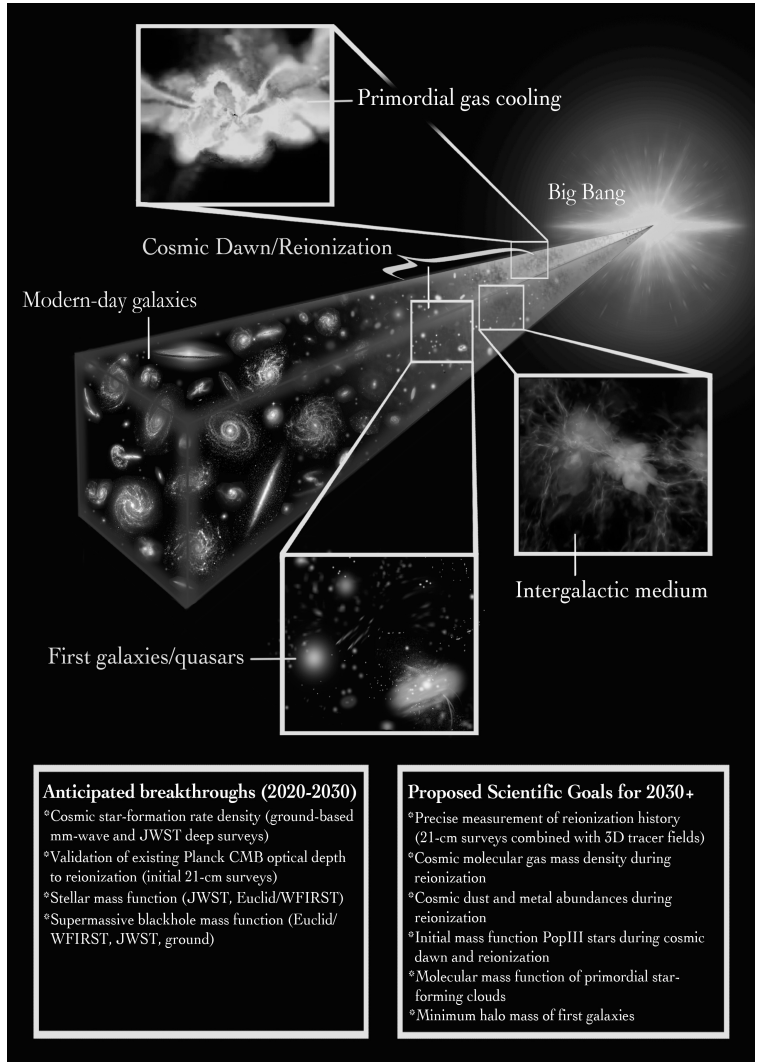

**Figure 1. Illustration of the cosmic dawn and reionization epoch in the history of the universe.** We also outline some of the key anticipated breakthroughs and proposed scientific goals for 2030 facilities.

**II. State of Play in 2019**

*Optical depth:* Although the basic framework is in place, we have significant deficiencies in our understanding of the physical processes at work during cosmic dawn and EoR. While the integrated optical depth is now known from Planck CMB polarization to a reasonable precision, the exact reionization history is largely uncertain with only information coming at the end of reionization with damping wings of $z \sim 6$-$7.5$ quasars (e.g., Davies et al. 2018), a small number of $z > 6$ Gamma-ray bursts (Salvaterra 2015), Ly-α emitters, Ly-α from drop-outs (Pentericci et al. 2014), and Ly-α absorption measurements.

*UV photon budget:* Our knowledge on the UV ionizing photon budget is limited by the small number of detected galaxies and quasars at $z > 6$, biases in the current sample selection, and the cosmic variance in pencil beam surveys. Most of the galaxy detections are either photometric drop-out selections using HST deep surveys complemented by ground-based data or narrow-band imaging to detect Ly-α emitters in atmospheric windows in the near infrared. As the detections are limited to mostly luminous galaxies, even with lensing magnification boost from galaxy clusters, statistical measurements such as the star-formation rate density (SFRD) and stellar mass density (SMD) during EoR require extrapolations of the measured LFs to fainter magnitudes. Despite detections of luminous quasars in the X-rays, cosmic black-hole accretion rate density (BHARD) remains largely uncertain at $z > 6$ as even deepest Chandra X-ray observations have yet to reach the depths necessary to detect typical AGNs during reionization.

*Escape fraction:* While galaxy samples exist at $z > 6$ we still lack observational measurements that directly connect UV photons to the surrounding IGM. We also lack information on the circumgalactic medium (CGM), gas accretion, and outflows. Our inferences on the escape fraction of the continuum UV and Ly-α photons are limited to indirect techniques such as stacking detected Ly-α galaxy locations on multi-band images leading to global averages. Numerical simulations are also inconclusive on the escape fraction.

*Dust and ISM:* Sub-mm and mm-wave surveys such as *Herschel* and SPT have allowed selections of luminous dusty starbursts at $z > 6$ (e.g., Riechers et al. 2013; Marrone et al. 2018). Such sources however are rare and do not represent typical galaxies that are needed to understand processes that were crucial during reionization. ALMA has now started to detect galaxies at z > 6, and as high as z=9.1 (Hashimoto et al. 2019), both in the dust continuum and fine structure emission lines such as [CII] 158 μm and [OIII] 88 μm. [OIII] detections strongly suggest the presence of young, possibly massive, stellar populations. Sub-mm spectral line searches with ALMA are likely the most effective avenue to spectroscopically confirm the brightest high-z galaxies. With observations limited to one galaxy at a time, observations of large samples adequate for statistical studies on the metal and dust production are still in its infancy. Since even a small amount of dust affects SEDs (e.g., Katz et al. 2019), we must expand dust studies to typical galaxies to improve our understanding of physical properties.

*IGM:* 21-cm studies that are capable of probing the ionization state of the IGM have only resulted in upper limits on the anisotropy power spectrum. The absorption feature in the global intensity spectrum seen by EDGES remains to be confirmed. Bright quasars have allowed IGM studies via the Ly-α absorption, allowing the end states of reionization be mapped towards selected lines of sights in the universe.

### III. Why study cosmic dawn and reionization with multiple probes?

A multi-wavelength strategy is crucial to study reionization since there are crucial links between different probes. For example, the heating of the IGM, stellar mass density built up during EoR, and feedback of metals to the IGM are key quantities that need to be measured since: (i) They drive/suppress the formation of low mass ($M_{halo} < 10^7$ $M_{sun}$) galaxies and globular clusters; (ii) The resultant metallicity and turbulence characterizes the stellar initial mass function at later times; and (iii) The nature of the stellar initial mass function (IMF) in the first galaxies provides insights into the formation of seed black holes.

### IV. What we expect in 2020-2030

*HST, existing, and new ground:* Deep surveys with HST, including approaches such as lensing frontier fields, and narrow-band surveys from the ground will continue to uncover z > 6 galaxies incrementally through photometric signatures. Existing and new instruments on ground-based optical observatories allow spectroscopy out to wavelengths < 2.4 μm, useful for science questions with detections of UV lines like Lyα and HeII 1640. Narrow-band imaging for Ly-a emitters can probe deeper in the LF, but will continue to be impacted by lack of redshift confirmations. Improvements to AO-assisted imaging and spectroscopy from ELT may resolve morphology and dynamics of luminous z > 6 galaxies to understand physical conditions. IFU spectroscopy with ELT will also allow detections of faint Ly-α emitters.

*JWST:* Deep pencil beam surveys with instruments on JWST will allow detections of fainter galaxies. JWST will open up the spectroscopic study of rest-frame optical lines at z > 6, allowing inferences on physical quantities such as the stellar mass, extinction, metallicity, and ionization state. JWST studies will be limited to a handful of narrow but deep fields. An ultra-deep survey with JWST – aided by cluster lensing - may be used to detect the faint-end turn-over expected in UV LFs. JWST spatial resolution will also aid resolved studies on reionizing galaxies.

*Euclid/WFIRST and LSST*: With their ability to carry out wide area surveys in near-IR from space, Euclid/WFIRST will improve the selection of z=6-10 galaxies and AGNs. Euclid, combined with LSST, will be able to select samples of z > 6 sources, though the shallow depth over its wide 15,000 deg$^2$ area limits the selection to brightest of the galaxies during EoR. The 40 deg$^2$ Euclid-deep fields will be a factor of six fainter and will improve the selection to lower luminosities. As photometric-selected Euclid/WFIRST samples will have large uncertainties on their exact redshifts. While Euclid/WFIRST surveys will lead to large samples of galaxies and AGN during reionization, their use will be limited due to lack of spectroscopic measurements (see Cuby et al. Science White Paper). Increased quasar samples from z=6-10 with Euclid/WFIRST will improve studies on reionization via damping wing measurements. The time overlap between JWST and Euclid/WFIRST is more useful since rare galaxies and bright quasars during the EOR can be followed-up for spectroscopic studies more efficiently with JWST. JWST alone however cannot follow-up all or even a reasonable fraction of the Euclid/WFIRST sources during EoR. Improvements to instruments on 10m class ground-based facilities will be needed to fully sample the z=6-10 quasar samples.

*SPHEREx:* All-sky spectroscopy with SPHEREx will allow excellent characterization of the bright-end of the quasar LF at z > 6 through rest-frame UV/optical spectroscopy out to an observed wavelength of 5 μm. These would again be ideal targets for HST, JWST, ALMA, and ELT.

*Ground and Space-based Intensity Mapping:* The extragalactic background light (EBL) fluctuation studies with CIBER, SPHEREx, etc should determine the total UV luminosity density during reionization. Such measurements are capable of addressing the photon budget fraction missed by detected galaxies. The intensity mapping results could be used to verify the importance of faint sources during EoR. Intensity mapping measurements can also guide deeper photometric surveys with JWST and WFIRST to resolve more of the photon budget to individual sources. Line intensity mapping focused on CO and CII etc from the ground at radio and mm wavelengths trace the collective emission from sources that are responsible for UV photons reionizing the universe. Such sources are anti-correlated with HI traced by the 21-cm line, allowing measurement of the HII bubble sizes. Second-generation intensity mapping experiments will need to match wide areas covered by 21-cm surveys (see Kovetz et al. Science White Paper).

**MM-wave:** For astrophysical applications the exact redshifts for such sources are crucial. For mm-wave spectroscopy, important rest-frame far-IR fine structure lines such as [CII] 158 μm and [OIII] 88 μm could be detected with ALMA, allowing limited studies related to star-formation and the ISM. ALMA redshift surveys could be improved with multi-beam feeds and bandwidth extensions. New instruments on LMT will improve the selection of typical galaxies at z > 6 with SFRs with 1-10 $M_{sun}$/yr, leading to additional galaxy samples for spectroscopic studies with ALMA to improve our understanding of ISM, metals and dust. With deeper data and careful sample selections, a combination of Euclid/WFIRST/JWST/mm-wave instruments/ALMA will allow SFRD and stellar mass density to be established at z > 6 during EoR, improving ten-folds over the current measurements (see Casey et al. Science White Paper).

*Chandra & X-rays:* With adequate galaxy samples selected from deep surveys with WFIRST and JWST, deeper Chandra X-ray observations and a combination of analysis techniques such as stacking may also allow first measurements of the BHAR during reionization. For a limited number of bright targets JWST may also studies relates to AGN activity through rest-frame optical spectroscopy in the mid-infrared.

*HERA, SKA-low, and 21-cm experiments:* Improvements to existing interferometers will likely reach the sensitivity level needed for first detections of the anisotropy power spectrum. Such measurements will be crucial to establish first detailed measurements on hydrogen reionization and IGM reheating. The history of reheating may not have coincided with the history of reionization. This possibility could be easily confirmed with 21-cm line intensity power spectrum measurements expected over the coming years with HERA, SKA-Low, Nenufar. The 21-cm anisotropy measurements will continue to be impacted by foregrounds and systematics. To fully extract the information content captured by 21-cm data it will become crucial to pursue cross-correlation studies with independent tracer fields during EoR (see Hutter et al. Science White Paper). Cross-correlations also allow an important to minimize foreground contaminations and systematics. WFIRST and Euclid+LSST, may provide sufficient galaxy samples for cross-correlation with 21-cm data, but the lack of precise redshifts limit their uses with photometric selections alone for cross-correlations.

## V. Science goals for 2030+ facilities

In light of the remarkable insights we will gain into first galaxies, AGNs, and reionization through facilities in the 2020s from 2000 (JWST) and 2010 (WFIRST/LSST) priorities, we recommend following scientific goals and objectives for new facilities that will begin science operations in 2030s:

**Science Recommendation #1: A precise measurement of the reionization and reheating history.** To claim a complete understanding of reionization we must first establish both the reionization and reheating histories, and their spatial variations. Cosmic-variance limited large angular scale CMB polarization allows an optical depth measurement with σ(τ)=0.002 vs 0.007 error in Planck. Improvements to 21-cm measurements will allow studies on both reionization and reheating history. 21-cm measurements reach full potential with ancillary tracer data that could be used for cross-correlation studies. Continued damping wing spectroscopy of z=6-10 bright to faint quasars selected from Euclid/WFIRST/SPHEREx etc with 10m to ELT class facilities from the ground will improve studies on reionization inhomogeneities.

**Technical Recommendation:** To allow maximal information extraction from 21-cm data, ground and space facilities capable of providing 3D tracer fields at z > 6, such as the galaxy density and Ly-α maps.

**Science Recommendation #2: Establish physical properties of galaxies and AGN during reionization.** Significant knowledge related to properties of galaxies and AGNs at moderate redshifts to today have come through spectroscopic observations in the optical and with statistics involving millions of galaxies (e.g., SDSS). Extending such studies to reionization then requires the ability to carry out large near- to mid-IR spectroscopic surveys over many 100-1000 deg$^2$ to reach sufficient cosmic volumes. Such rest-frame optical spectroscopic measurements will be capable of establishing SFRs, extinction, metallicity, IMF, ionization state, AGN accretion, and the presence of PopII and PopIII stars in first-light galaxies.

Galaxy properties are also best studied in the rest-frame mid to far-infrared emission lines, extending limited approaches at low redshifts with *Spitzer*/IRS and *Herschel*, and bridging the wavelength gap between JWST and ALMA. To be effective for reionization studies far-infrared observations will need to reach the depths sufficient to detect typical galaxies at $z > 6$. The collecting area and sensitivity of SPICA is such that spectroscopic studies at $z > 2$ will be primarily limited to IR luminous, dust-rich starburst galaxies – at $z > 6$ even those rare luminous galaxies will be undetectable. The primary goals of a new mission, with a factor of 20-30 higher sensitivity than SPICA, would be then to acquire physical properties of typical $z > 6$ galaxies leading to studies on the physical conditions of star-formation and ISM, IMF, rise of metals, formation mechanisms of dust, central blackhole accretion, among others and all unaffected by dust attenuation that impacts rest-frame optical lines (see Pope et al. Science White Paper).

**Technical Recommendation:** Mid IR and far-IR space-based facility capable of carrying out spectroscopic studies during reionization, leading to statistically adequate samples of galaxies and AGNs at $z > 6$.

**Science Recommendation #3: Study Lyman-α to understand the connection between sources and the IGM.** Because of radiative transfer effects, the observed Lyα luminosity of a galaxy is lower than its intrinsic luminosity. This difference is characterized by the Lyα escape fraction. The global Lyα escape fraction can be estimated by the ratio of Lyα and Hα/UV luminosity densities, though to understand reionization we need to establish both global or average Lyα escape fraction as well as its spatial and/or environmental dependences. While at low redshifts, the observed Lyα escape fraction is mainly due to radiative transfer in the ISM and CGM (e.g. dust absorption and spatial diffusion), during reionization, IGM transmission is expected to play a major role in shaping the Lyα escape fraction. A combination of 21-cm measurements, galaxy properties, and maps of the extended and diffuse Lyα emission during reionization can be used to connect IGM to sources of photons (See Chang et al. Science White Paper).

**Technical Recommendations:** A space-based facility capable of 3D Lyman-α imaging over large volumes at $6 < z < 15$ and with sensitivity to the diffuse IGM emission.

**Science Recommendation #4: Study the formation of seed blackholes.** SDSS and other luminous quasars at $z > 6$ have estimated blackhole masses in excess of billion solar masses. While such super-massive blackholes could have been seeded by stellar remnants of first stars (with masses out to 100 solar masses), they must accrete at super-Eddington rates for most of their lifetime (Pezzulli et al. 2017). The other option is seeds that are $10^4$ $M_{sun}$, resulting from direct collapse of gas in protogalactic disks. LISA gravitational wave observations would be capable of distinguishing between these two options to disentangle seeding channels from accretion physics (Ricarte & Natarajan 2018). Studies on the abundance and spatial distribution of seed blackholes require deep X-ray observations with Athena in the 2030s. Athena or a new X-ray facility may also allow for rapid follow-up of LISA events (see Natarajan et al. Science White Paper).

**Technical Recommendation:** X-ray facility with sensitivity, wide area mapping, and LISA counterpart/transient identification capability for $z > 6$ sciences.

**Science Recommendation #5: Study the formation epoch of first sources during cosmic dawn.** JWST will likely be capable of detecting galaxies at the onset of reionization, when such galaxies have formed a sufficient number of stars and rest-frame UV/optical bright. JWST observations, however, are not adequate to address the formation epoch of the galaxies and the minimum mass of first-light galaxies. For such studies we must pursue observations even earlier during cosmic dawn to study the cooling of primordial gas. As metals are not present, existing theoretical models suggest that cooling is regulated by molecular hydrogen. The primary cooling radiation appears in the mid-IR $H_2$ vibro-rotational lines, with dominant emission lines in the rest wavelengths between 7 to 28 microns (see Appleton et al. Science White Paper).

**Technical Recommendation:** A far-infrared facility with sufficient sensitivity to detect primordial gas cooling at $10 < z < 20$ (perhaps aided by lensing magnification).

## References

Appleton, P. et al. 2019, *Warm H2 in Turbulence and Shock across Cosmic Time: From Local Galaxies to the Dark Ages*, Science White Paper to Astro2020

Barkana, R., Loeb, A., 2001. In the beginning: the first sources of light and the reionization of the universe. Physics reports. 349, 125-238.

Bouwens, R., et al., 2015. UV luminosity functions at redshifts z~ 4 to z~ 10: 10,000 galaxies from HST legacy fields. The Astrophysical Journal. 803, 34.

Bowman, J. D. et al. 2018, An absorption profile centred at 78 megahertz in the sky-averaged spectrum, Nature, 555, 7694, 67

Bromm, V., Larson, R. B., 2004. The first stars. Annu. Rev. Astron. Astrophys. 42, 79-118.

Casey, C. et al. 2019, *Taking Census of Massive, Star-Forming Galaxies formed <1 Gyr After the Big Bang*, Science White Paper to Astro2020

Chang, T-C. et al. 2019, *Tomography of the Cosmic Dawn and Reionization eras with multiple tracers*, Science White Paper to Astro2020

Chary, R., Petitjean, P., Robertson, B., Trenti, M., Vangioni, E., 2016. Gamma-ray bursts and the early star formation history. Space Science Reviews. 202, 181-194.

Cuby, J.G. et al. 2019, *Understanding Reionization: the key role of ground-based telescopes*, Science White Paper to Astro2020

Davies, F. B. et al. 2018, Quantitative Constraints on the Reionization History from the IGM Damping Wing Signature in Two Quasars at z > 7. The Astrophysical Journal. 864, 142.

Dayal, P. & Ferrara, A. 2018, Early galaxy formation and its large-scale effects, Physics Reports, 780, 1

Finkelstein, S. L., et al., 2015. The evolution of the galaxy rest-frame ultraviolet luminosity function over the first two billion years. The Astrophysical Journal. 810, 71.

Finkelstein, S. L. et al. 2019, Conditions for Reionizing the Universe with A Low Galaxy Ionizing Photon Escape Fraction, arXiv.org:1902.02792

Furlanetto, S. & Loeb, A. 2004, Large-Scale Structure Shocks at Low and High Redshifts. The Astrophysical Journal. 611, 642

Hashimoto, T. et al. 2018, The onset of star formation 250 million years after the Big Bang, Nature, 557, 392

Hayes, M., Schaerer, D., Östlin, G., Mas-Hesse, J. M., Atek, H., Kunth, D., 2011. On the redshift evolution of the lyα escape fraction and the dust content of galaxies. The Astrophysical Journal. 730, 8.

Hutter, A., et al. 2019, *A proposal to exploit WFIRST-SKA synergies to shed light on the Epoch of Reionization*, Science White Paper to Astro2020

Katz, H. et al. 2019, Probing cosmic dawn: modelling the assembly history, SEDs, and dust content of selected z ~ 9 galaxies. Monthly Notices of the Royal Astronomical Society. 484, 4054-4068.

Kovetz, E. et al. 2019, *Astrophysics and Cosmology with Line-Intensity Mapping*, Science White Paper to Astro2020

Johnson, J. L. & Khochfar, S. 2011, The Contribution of Supernovae to Cosmic Reionization. The Astrophysical Journal. 743, 2

Madau, P., Haardt, F., 2015. Cosmic reionization after planck: Could quasars do it all? The Astrophysical Journal Letters. 813, L8.

Madau, P., Meiksin, A. & Rees, M. J. 1996, 21-cm Tomography of the Intergalactic Medium at High Redshift. The Astrophysical Journal. 475, 429

Mapelli, M., Ferrara, A. & Pierpaoli, E. 2006, Impact of dark matter decays and annihilations on Reionization. Monthly Notices of the Royal Astronomical Society. 369, 1719

Marrone, D. et al. 2018, Galaxy growth in a massive halo in the first billion years of cosmic history, Nature, 553, 51

Natarajan, P. et al. 2019, *Disentangling nature from nature: understanding the origin of seed black holes*, Science White Paper to Astro2020
- 7 -